# Physics-Informed Graph Neural Network for Inverse Design of Integrated Photonic Biosensors

Yasaman Torabi[1,*], Amirali Ekhteraei[2], Mohammad Khajezadeh[1]
[1]Department of Electrical and Computer Engineering, McMaster University, Hamilton, ON L8S 4L7, Canada
[2]Department of Physics, McGill University, Montreal, QC H3A 2T8, Canada
[*]Corresponding author: torabiy@mcmaster.ca

*Abstract*— Integrated photonic biosensors provide compact, highly sensitive, and label-free platforms for biochemical detection, making them attractive for on-chip and real-time sensing applications. However, their design remains challenging due to complex resonance behaviour, strong coupling effects, and the computational cost associated with repeated full-wave electromagnetic simulations. In particular, inverse design of microring resonator-based sensors requires accurate modeling of geometry–spectrum relationships while satisfying physical constraints such as resonance conditions and spectral sensitivity requirements. In this work, we propose a physics-informed graph neural network (PI-GNN) framework for the inverse design of a microring resonator biosensor operating in the 1550 nm band. By representing the photonic structure as a graph and embedding resonance-based physical constraints directly into the learning objective, the model captures both structural connectivity and underlying electromagnetic principles. The proposed approach enables efficient prediction of device geometries that achieve target spectral characteristics, reducing reliance on costly simulations while maintaining physical consistency and competitive design accuracy.[1]

**Keywords**— AI for photonics, integrated photonics, inverse design, graph neural network, physics-informed neural network, biosensor

## Introduction

Integrated photonic biosensors provide compact and highly sensitive solutions for label-free biochemical detection, enabling real-time monitoring of refractive index variations caused by molecular binding events. Among these devices, microring resonator-based sensors are widely used due to their high quality factors, small footprint, and compatibility with standard fabrication processes. Their sensing performance strongly depends on precise control of geometric parameters, optical coupling strength, and material properties. Small variations in ring radius, waveguide width, thickness, or coupling gap can significantly shift the resonance wavelength and alter spectral sensitivity, making accurate design both critical and challenging [1]. Conventional design approaches rely on finite-difference time-domain (FDTD) simulations, eigenmode analysis, or adjoint-based optimization. Although these techniques provide accurate electromagnetic modeling, they are computationally intensive and limit efficient exploration of high-dimensional design spaces. Inverse design, where target spectral characteristics are specified and the corresponding geometry is inferred, is particularly difficult due to nonlinear resonance behavior, parameter coupling, and the presence of multiple feasible solutions. Consequently, reducing simulation cost while maintaining physical accuracy remains an important objective in photonic device design.

Recent advances in artificial intelligence (AI) have introduced data-driven strategies for accelerating photonic inverse design. Variational autoencoders (VAEs) [2], diffusion models (DMs) [3], and reinforcement learning (RL) [4] have demonstrated the ability to approximate complex geometry–response relationships and reduce dependence on repeated full-wave simulations. Graph neural networks (GNNs) [5,6] provide an additional advantage by representing devices as interconnected structures composed of nodes and edges. This representation aligns naturally with photonic components such as rings, bus waveguides, coupling regions, and surrounding media, allowing structural relationships to be encoded explicitly. Unlike conventional neural networks that treat device parameters as independent inputs, GNNs capture relational dependencies between components. This property enables the integration of physical constraints directly into the learning process. Graph connectivity encodes physical structure and makes the model "physics-aware", while it becomes "physics-informed" when governing physical relationships are explicitly enforced in the loss function during training. Physics-informed graph neural networks (PI-GNNs) have shown strong performance in electromagnetic modeling and related physics-driven tasks [5], and recent AI-based photonic design studies further highlight the importance of embedding physical knowledge into machine learning architectures such as graph neural networks for better sensor fabrication [7]–[9]. However, their application to inverse design of microring-based biosensors under explicit resonance and spectral sensitivity constraints has not been systematically investigated. In this work, we introduce a physics-informed graph neural network for inverse design of integrated photonic biosensors. The model incorporates the microring resonance condition and spectral sensitivity requirements into the loss function, ensuring consistency between predicted geometries and electromagnetic behavior. By combining graph-based structural representation with resonance-based physical constraints, the proposed method improves inverse design stability, reduces computational burden, and provides physically consistent solutions for targeted sensing performance.

---

[1] The Python scripts are publicly available on GitHub at: https://github.com/Torabiy/PIGNNRing

## Methodology

***Biosensing Mechanism:*** Microring resonators enable biochemical sensing by tracking resonance shifts induced by refractive-index (RI) changes after biomolecular bindings (Fig. 1). Prior studies demonstrate that low-index polystyrene polymer microring waveguides ($n_{eff} \approx 1.53$) achieve acceptable sensitivity [10]. Microring sensors with a quality factor of $Q \approx 2\times10^4$ achieve detection limits of $10^{-5}$ RIU, desirable for label-free biosensing.

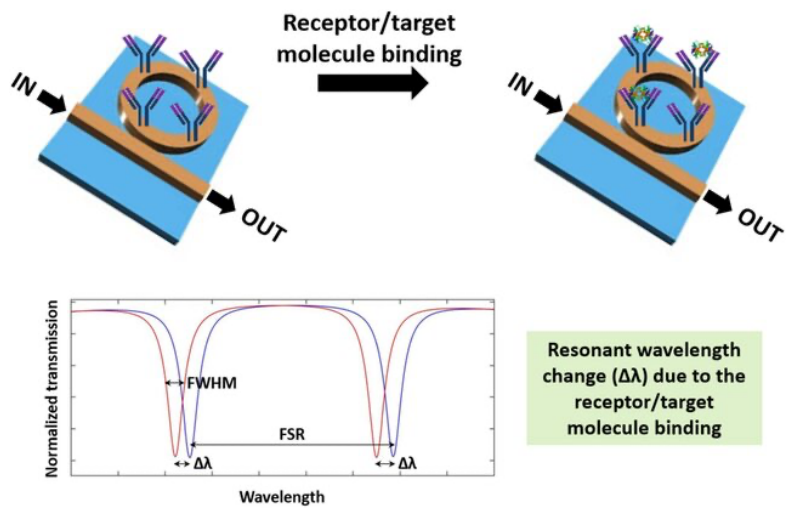

**Fig. 1** Sensing mechanism of an optical ring resonator biosensor [11]

The sensing principle is governed by the resonance condition requiring that the accumulated round-trip phase equals an integer multiple of $2\pi$. For a ring of radius R, the round-trip length is L=$2\pi$R, and resonance occurs when $\beta L = 2\pi m$, where $\beta = \frac{2\pi}{\lambda} n_{\text{eff}}$ is the propagation constant, $\lambda$ is the vacuum wavelength, $n_{eff}$ is the modal effective index, and $m \in \mathbb{Z}$ is the azimuthal resonance order. Substituting $\beta$ yields $m\lambda_0 = n_{eff}2\pi R$, which explicitly links the resonance wavelength $\lambda_0$ to both geometry (through $R$) and modal properties (through $n_{eff}$). When biomolecular binding induces a small cladding refractive index perturbation $\Delta n_{clad}$, the effective index changes by $\Delta n_{eff}$. Differentiating the resonance condition and accounting for dispersion leads to the first-order wavelength shift $\Delta\lambda = \frac{\lambda_0}{n_g}\Delta n_{eff}$, where $n_g$ is the group index. Using electromagnetic perturbation theory, the effective index variation is proportional to the modal overlap with the sensing region, $\Delta n_{eff}=\Gamma\Delta n_{clad}$, where the confinement factor $\Gamma$ represents the fraction of modal energy interacting with the analyte. Combining these relations yields the bulk sensitivity

$$S = \frac{\Delta\lambda}{\Delta n_{clad}} = \frac{\lambda_0}{n_g}\Gamma.$$

This expression clarifies that sensitivity increases with stronger evanescent field overlap $\Gamma$ and decreases with larger group index $n_g$, revealing the geometric–modal trade-offs inherent in microring design. The resonance sharpness is characterized by the loaded quality factor

$$Q = \frac{\lambda_0}{\Delta\lambda_{FWHM}},$$

where $\Delta\lambda_{FWHM}$ denotes the full width at half maximum of the resonance dip. In terms of total propagation loss coefficient $\alpha_{tot}$, the intrinsic quality factor can be written as

$$Q_{int} = \frac{2\pi n_g}{\alpha_{tot}\lambda_0}.$$

Finally, the limit of detection (LOD) is determined by

$$\text{LOD} = \frac{\delta\lambda_{min}}{S},$$

where $\delta\lambda_{min}$ is the minimum resolvable wavelength shift limited by system noise and spectral resolution.

In addition to sensitivity and quality factor, the free spectral range (FSR), defined as the wavelength spacing between adjacent resonances, plays an important role in practical sensing operation. The FSR is primarily determined by the ring radius and the group index of the guided mode. Larger radii lead to smaller FSR values, resulting in more closely spaced resonances, while smaller radii increase the FSR and provide wider spectral separation between modes. A sufficiently large FSR is desirable to avoid spectral overlap and ensure unambiguous tracking of resonance shifts during sensing measurements. However, reducing the ring radius to increase FSR may also influence bending losses and modal confinement, introducing additional trade-offs in design. These coupled relations demonstrate that small geometric variations in $R$, waveguide width $w$, height $h$, and coupling gap $g$ simultaneously influence $n_{eff}$, $n_g$, confinement factor $\Gamma$, propagation loss $\alpha_{tot}$, and therefore the resonance wavelength, sensitivity, and detection limit. This strong nonlinear dependency motivates incorporating resonance-consistent physical constraints into the inverse design process to ensure that predicted geometries remain electromagnetically valid while achieving target sensing performance.

*Optical Design Consideration:* We define the sensor using the ring radius *R*, waveguide width *w*, height *h*, and coupling gap *g*, and we vary these parameters over 0–500 μm, 1.5–3.0 μm, 1.5–3.0 μm, and 0.1–0.3 μm, respectively. The selected parameter ranges are chosen to balance practical fabrication feasibility with sufficient design diversity for learning. The ring radius controls the optical path length and therefore directly influences the resonance wavelength and free spectral range. The waveguide width and height determine modal confinement, effective index, and bending loss, while the coupling gap governs the interaction strength between the bus waveguide and the ring, affecting extinction ratio and loaded quality factor. Because these parameters are strongly coupled, variations in one dimension can indirectly alter spectral response through changes in effective index, group index, and propagation loss. We generate two datasets from the analytical model based on coupled-mode theory and the FDTD simulations. The analytical dataset provides fast evaluations based on resonance and coupling relations, enabling efficient sampling of the design space. In contrast, the FDTD dataset captures full electromagnetic field interactions, including higher-order effects, bending-induced loss, and coupling non-idealities. By combining both data sources, we ensure that the model learns both physically structured trends from theory and high-fidelity behavior from numerical simulations. The device operates in the 1550 nm wavelength band. We select the resonance dip closest to 1.55 μm, and record the baseline resonance wavelength $\lambda_0$ together with its spectral shift $\Delta\lambda$ under varying RI conditions.

*Physics-Informed Graph Neural Network*: A PI-GNN incorporates physical structure into machine learning. We represent the sensor as a graph whose nodes correspond to the ring, bus waveguide, coupling region, surrounding medium, and substrate. Edges encode physical interactions between these regions, such as waveguide coupling and evanescent field interaction (Fig. 2).

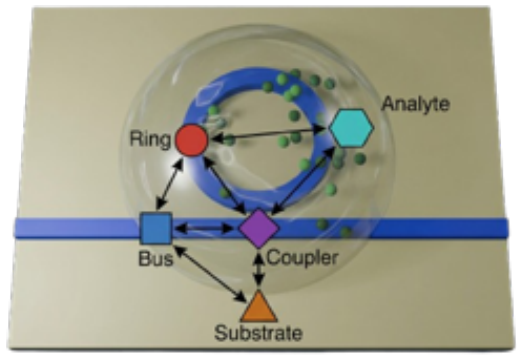

**Fig. 2** Graph abstraction of the microring biosensor. The physical device is mapped onto a structured graph in which nodes represent functional regions, including the microring, bus waveguide, coupling region, surrounding analyte medium, and substrate. Edges describe physical interactions such as optical coupling, field confinement, and evanescent overlap between adjacent components. This graph representation enables message passing across interconnected regions, allowing the model to learn how local geometric variations collectively determine global spectral response.

Each node is assigned feature vectors describing relevant geometric and material parameters, including local dimensions, refractive indices, and region-specific properties. Edge features characterize interaction mechanisms such as coupling strength, field overlap, and adjacency relationships. This structured representation enables the model to propagate information across physically connected regions through message-passing operations. Unlike conventional fully connected networks that treat geometric parameters independently, the graph formulation preserves relational dependencies and ensures that learning respects device connectivity. To further incorporate physical consistency, the learning objective integrates resonance-based constraints. This guides the network toward solutions that not only match target spectral outputs but also remain electromagnetically valid. By combining graph-based structural encoding with physics-informed loss terms, the PI-GNN captures nonlinear geometry–spectrum relationships while reducing reliance on repeated full-wave simulations. This approach enhances stability during inverse optimization and improves generalization across the explored design space.

We train the network to learn the relationship between device geometry and spectral response (Fig. 3a). During the forward stage, the network receives geometric parameters associated with each node and propagates information across the graph through iterative message-passing steps. At each layer, neighboring nodes exchange encoded representations, allowing local geometric variations in the ring, coupling region, and bus waveguide to collectively influence the global spectral prediction. The aggregation and pooling operations transform node-level information into a compact device-level embedding, which is then mapped to spectral quantities such as resonance wavelength and sensitivity-related metrics. This process enables the model to approximate the nonlinear mapping from geometry space to spectral response without explicitly solving Maxwell's equations at each iteration. After training, we use the learned model to identify geometries that satisfy target spectral features (Fig. 3b). In the inverse stage, the trained network is embedded within an optimization loop. Instead of directly predicting spectral outputs, the objective is reversed: desired spectral characteristics are specified, and the model searches for geometric parameters that minimize the discrepancy between predicted and target responses. The data loss term enforces agreement with target spectral quantities, while the physics-informed loss ensures that predicted resonance behavior remains consistent with the microring phase condition. This dual-objective formulation stabilizes the inverse design process and reduces the likelihood of physically inconsistent solutions.

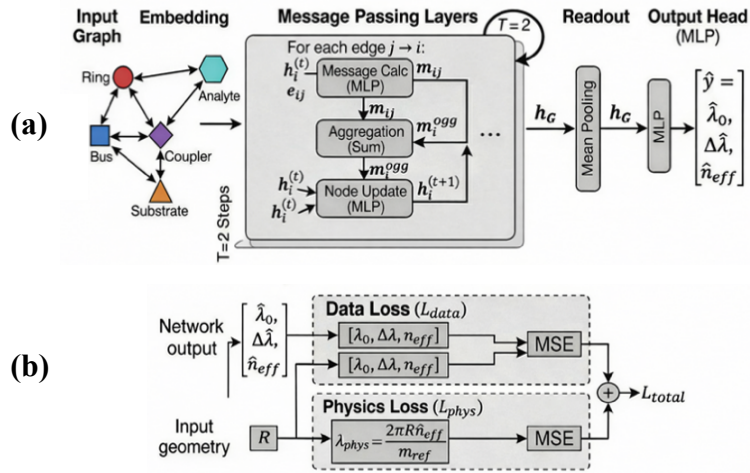

**Fig. 3** Physics-Informed Graph Neural Network (PI-GNN). **(a)** Forward stage, consisting of a linear node embedding layer, two message-passing layers (T=2), sum aggregation, global mean pooling, and a multilayer perceptron (MLP) output head. Here, $h_i$ denotes the hidden feature vector of node $i$, $e_{ij}$ is the edge feature between nodes $i$ and $j$, and $m_i$ is the aggregated message from neighboring nodes. **(b)** Inverse stage, combining data loss on predicted quantities with a physics-informed loss. The upper mean square error (MSE) block enforces agreement between predicted and target spectral quantities, while the lower MSE enforces consistency between the predicted resonance wavelength $\lambda_0$ and the microring resonance condition. The value $m_{ref}$ denotes the resonance order, fixed at the target geometry.

## Results

We benchmarked against a previously tested design in [9], and we evaluated geometric deviation from the target design ($R$=200 µm, $w$=2 µm, $h$=2 µm, and $g$=0.25 µm) by targeting $\lambda_0 \approx$1.55 µm and $\Delta\lambda < 0.5$ nm. The dataset consists of 2,500 geometries (80/20 split). We trained for 80 epochs and validated 50 designs using FDTD. The inverse design converged to $R$=196.620 µm, $w$=2.054 µm, $h$=1.973 µm, and $g$=0.226 µm, yielding a resonance wavelength $\lambda_0$=1550.009 nm and a spectral shift $\Delta\lambda$=0.1602 nm. We define the Root-Mean-Square Relative Geometric Error (RMS-RGE) as the square root of the mean squared relative deviations of the geometry parameters from their target values. The proposed model achieves RMS-RGE=5.14% (Analytical) and 3.46% (FDTD). Under identical settings, PI-GNN achieves lower overall error than DM [3], while the DM outperforms in generating the coupling gap $g$ (Fig. 3).

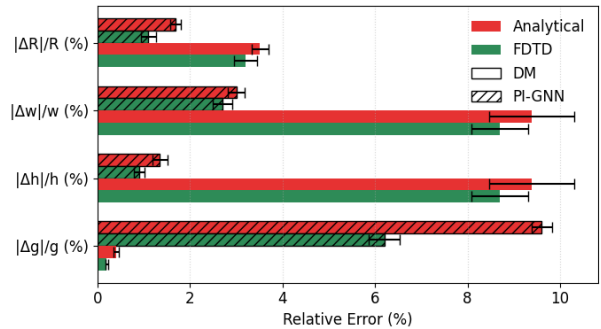

**Fig. 3** Comparison of diffusion model (DM) and physics-informed graph neural network (PI-GNN) performance across individual geometric parameters using analytical and FDTD datasets. Bars represent relative errors for ring radius, waveguide width, height, and coupling gap. PI-GNN demonstrates lower aggregate error across most parameters, particularly when trained on FDTD data, while DM shows competitive performance for the coupling gap.

The improved stability of PI-GNN arises from embedding resonance-consistent physical constraints directly into the graph structure and loss formulation, which guides the optimization toward electromagnetically valid solutions and reduces the need for repeated FDTD evaluations during training. As shown in Fig. 4, PI-GNN demonstrates stable and monotonic convergence behavior across both analytical and FDTD datasets, with training and testing curves following consistent trends. An ablation study comparing PI-GNN with an identical GNN architecture without the physics-informed term shows that PI-GNN achieves lower test error on both analytical and FDTD datasets. Although the GNN achieves lower training loss, its higher test error reflects weaker generalization and reduced adherence to the governing resonance physics.

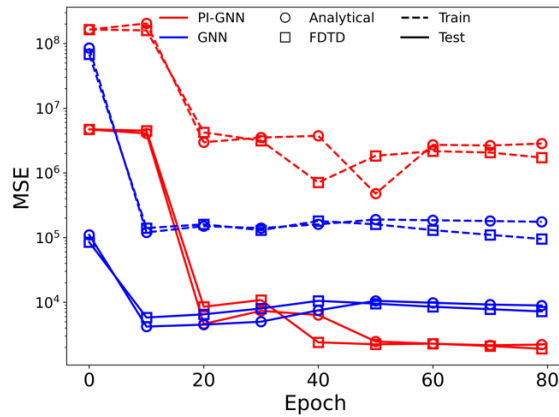

**Fig. 4** Ablation study of PI-GNN versus an identical GNN without the physics-informed term. Training and testing learning curves for the analytical and FDTD datasets are presented.

The rapid decrease in mean squared error (MSE) during early epochs indicates efficient learning of the dominant geometry–spectrum relationships, while later epochs show gradual refinement and stabilization. Notably, the analytical dataset exhibits higher steady-state error compared to the FDTD dataset. This difference suggests that simplified analytical models may introduce approximation errors that limit achievable accuracy, whereas the FDTD-based dataset provides higher-fidelity electromagnetic information that enables more precise learning. Despite this discrepancy, convergence remains smooth for both cases, confirming the robustness of the physics-informed architecture. The absence of significant oscillations or divergence in the learning curves highlights the stabilizing effect of incorporating physical constraints into the graph-based learning process. Moreover, the PI-GNN performs competitively with other methods using their respective benchmarks (Table 1).

Table 1. Comparison of AI Inverse-Design Methods

| Ref | AI Model | Performance | Application |
|---|---|---|---|
| [2] | VAE | Up to 95% Spectral Accuracy | Silicon photonics |
| [3] | DM | Chamfer 0.03 (FDTD), 0.04 (Analytical) | Biosensing |
| [4] | RL | Dispersion +41.8% over PSO | Metasurfaces |
| [5] | PI-GNN | RMSE ≈ 2.21%, $R^2$ = 0.992 | 4D flow MRI |
| This Work | PI-GNN | RMS-RGE 3.46% (FDTD), 5.14% (Analytical) | Biosensing |

## Conclusion

In this work, we presented a physics-informed graph neural network for the inverse design of microring resonator-based integrated photonic biosensors. By encoding device topology as a graph and embedding resonance-consistent physical constraints into the learning objective, the proposed approach enables stable and accurate mapping between geometric parameters and spectral response. The model demonstrates competitive geometric reconstruction accuracy compared to diffusion-based methods while exhibiting improved convergence stability and reduced reliance on repeated full-wave simulations. Results obtained from both analytical and FDTD datasets confirm that incorporating physical structure into the learning process enhances robustness and predictive reliability across the explored design space. Despite these promising results, the current model does not yet incorporate biochemical selectivity or fabrication constraints. Future work will integrate fabrication-aware constraints and antibody–antigen modeling to enhance practical biosensing reliability and move toward experimentally realizable, application-ready photonic sensor designs.